# Advancing Tumor Budding Detection with Fourier Ptychography Microscopy

**Author names**

Yiyan Su[1, †], Ruiqing Sun[2, †], Yajing Wang[3], Yanfeng xi[3, *], Shaohui Zhang[4, *]

**Affiliations**

[1] Department of Health Statistics, School of Public Health, Shanxi Medical University, 56 South XinJian Road, Taiyuan 030001, Shanxi Province, China.

[2] School of Computer Science and Technology, Beijing Institute of Technology, Beijing, China

[3] Department of Pathology, Shanxi Province Cancer Hospital/ Shanxi Hospital Affiliated to Cancer Hospital, Chinese Academy of Medical Sciences/Cancer Hospital Affiliated to Shanxi Medical University, Taiyuan, China.

[4] School of Optics and Photonics, Beijing Institute of Technology, Beijing 100081, China

[†]Both authors contributed equally to the paper

**Corresponding author information**

Yanfeng Xi, Department of Pathology, Shanxi Province Cancer Hospital/ Shanxi Hospital Affiliated to Cancer Hospital, Chinese Academy of Medical Sciences/Cancer Hospital Affiliated to Shanxi Medical University, Taiyuan, Shanxi, People's Republic of China, 030000, Email: xiyanfeng1998@163.com

Shaohui Zhang, School of Optics and Photonics, Beijing Institute of Technology, Beijing, People's republic of China, 100081, Email: zhangshaohui@bit.edu.cn

**ABSTRACT**

**Background:** Tumour budding is an independent predictor of metastasis and prognosis in colorectal cancer and is a vital part of the pathology specification report. In a conventional pathological section observation process, pathologists have to repeatedly switch from 10x objective to 20x objective several times to localize and image the target region. Besides the switching operations, repeated manual or electro-mechanical focusing is also very time-consuming, affecting the total time for pathological diagnosis. In addition, It is usually necessary to remove the manually marked symbols on the stained pathology slides used for classification and management before observation.

**Methods:** In this paper, we utilize Fourier ptychographic microscopy (FPM) in the pathological diagnosis process to realize large space-bandwidth product imaging, quantitative phase imaging, and digital refocusing in the observation process without any mechanical operations, which can therefore simplify the above-mentioned cumbersome diagnostic processes. We first verify the effectiveness and efficiency of the proposed method with several typical pathological sections. Then, instead of manually erasing, we also prove that FP framework can digitally remove the artificial markers with its digital refocusing ability.

**Results:** At last, we demonstrated pathologists can achieve 100% diagnostic accuracy with FPM imaging results.

**Conclusions:** The proposed method can greatly simplify the process of pathological diagnosis, and the related addon hardware system does not require expensive components, which makes it have great potential for promotion in the field of pathological diagnosis.

1. **Introduction**

Currently, cancer is a serious disease that threatens the survival and death of human beings around the world. According to the statistics of 2022, there are more than 1.9 million new cases of colorectal cancer, which is the third most frequently diagnosed malignant tumor in the world, and the second leading cause of cancer death. Moreover, the incidence of cancer is trending younger, with an annual increase rate of 1-4%.[1]. Tumor budding is considered a significant independent prognostic indicator for colorectal cancer [2], defined as the presence of single cells or clusters of up to four cells scattered within the stroma at the invasive front of colorectal cancer. It is an independent factor for metastasis and prognosis in stage I and II colorectal cancer and is specified as part of routine reporting of colorectal cancer [3-7]. At the same time, the grading of tumor budding can help pathologists to provide an effective reference for the risk of lymph node metastasis, neoadjuvant therapy and surgical plan of the patient. As a valuable prognostic factor, it is increasingly being paid attention to by gastrointestinal pathologists [8-9]. The clinical and biological significance of tumor budding has developed profoundly in colorectal cancer and some other cancers [10-12].

In general, the budding sites tend to show clusters of cells with large heterogeneity, and because of the strong inflammatory response at the most anterior margin of the tumor cell infiltration, there are large numbers of inflammatory cells and neoplastic clusters that interfere with the pathologist's interpretation. In fact, clinical pathologists always have to combine haematoxylin-eosin (HE) staining and immunohistochemical staining (IHC) sections together to avoid interference and therefore make a more accurate diagnosis [13-16]. In a specific operation step, the pathologists first execute a precisely manual focusing procedure with a 10x objective to identify the area of tumor budding within the HE and IHC stained sections. Secondly, they switch to a 20x objective and perform more precise and carefully refocusing to locate the enumeration of the budding foci. Then, the pathologists revert to the 10x objective, and perform a fine-tuning of the focus, ensuring optimal visualization for the assessment. The above processes are always repeated several times, and finally the area with the highest number of outgrowths can be selected for counting and grading. Meanwhile, with the growing incidence of colorectal cancer, manually assessing tumor budding with a standard conventional microscope has become increasingly tedious, escalating the clinical workload. Furthermore, stained pathology slides, marked with symbols and stains for classification and management, can obscure the budding areas and tumor cells, complicating diagnostic accuracy. This highlights the need for an alternative to conventional microscopy in interpreting tumor budding.

As a typical computational microscopy, Fourier ptychographic microscopy (FPM) [17-18] can provide a large field of view and high spatial resolution simultaneously. And it has automatic refocusing and cleaning functions for the sections, which means that manual adjustment of focus is not required, nor is the need for manual wiping of dust, stains, and artificial marking symbols. What's more, unlike traditional microscopes, FPM can simultaneously obtain intensity and quantitative phase images during the imaging process, providing additional useful references for diagnosis [19-20]. For instance, Marika Valentino and colleagues used phase information to identify renal tissues without the need for staining [21]. Vittorio Bianco and his team used images obtained with FPM to differentiate between breast cancer and fibroadenomas of the breast [22]. Leveraging its digital refocusing capabilities, Anthony Williams and colleagues also achieved the counting of circulating tumor cells on different focal planes [23]. Even though, the aforementioned work still does not fully exploit the potential of FPM in the diagnosis of cancer.

In this paper, different from prior studies, we delve deeply into the advantages of FPM and apply them to real-world clinical pathological slides observation process. With its large field-of-view, high-resolution, digitally refocused imaging capability, it enables the identification of tumor budding hotspots in colorectal cancer and the counting of budding without the tedious switching between high and low NA objectives. Currently, the most fundamental approach to detect tumor budding is through Hematoxylin-Eosin (HE) staining, as recommended by the International Tumor Budding Consensus Conference (ITBCC). In this method, the pathologist examines the stained sections of the tumor and selects the one that shows the highest number of buddings at the invasive front. [3]. During the observation of tumor budding, we only utilize a 10x objective, without the need for additional knob turning for focusing or objective lens switching. On this basis, by combining HE and IHE sections with each other, the pathologists can avoid interference and then make a better diagnosis decision [2]. In fact, no matter whether relying on He or IHE, both approaches require several repeatable switches between low-NA and high-NA objectives and mechanical refocusing. In addition, we have implemented an FPM system and utilized it in a real pathological diagnosis. We conducted real comparison experiments to demonstrate the feasibility and effectiveness of the proposed tumor budding detection method. Therefore, compared with the conventional microscope and related observation and diagnosis procedures, our method has the characteristics of low cost, ease of implementation and excellent performance, making it suitable for application to a large number of clinical works, and therefore can be widely spread in the near future. Next, we will compare the images obtained by FPM, with its unique advantages, to those from traditional microscopes to demonstrate the above-mentioned points.

**2. Materials and methods**

By introducing FPM, a typical computational microscopy framework, we hope to simplify the process of tumor budding detection for pathologists who need to repeatedly switch objective lenses and mechanically refocus. Meanwhile, the function of FPM to automatically remove contaminants on the section surface further optimises the tumor microscopy diagnostic process. The specific FPM diagnostic flow is shown in Figure 1.

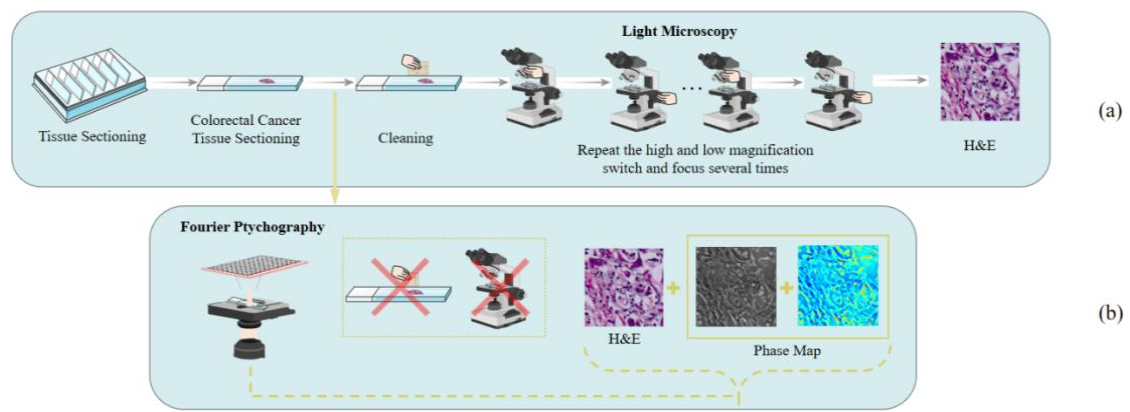

Fig. 1 Assessing tumor budding tissue sections: (a)Analysis with a conventional optical microscope. (b)Analysis with Fourier ptychographic microscopy.

2.1 Slides preparation

To implement and validate the effect of FPM in tumor budding detection, we randomly selected and analyzed 600 cases of colorectal cancer with tumor budding, reported over the past three years. These cases included specimens with both immunohistochemistry (IHC) and hematoxylin-eosin (HE) staining. Experienced pathologists reviewed the cases using both conventional microscopy and Fourier ptychographic microscopy (FPM) to assess the consistency of diagnostic outcomes between the two techniques. The aim was to show that FPM can achieve the same diagnostic results while simplifying the clinical workflow.

2.2 FPM principle and workflow

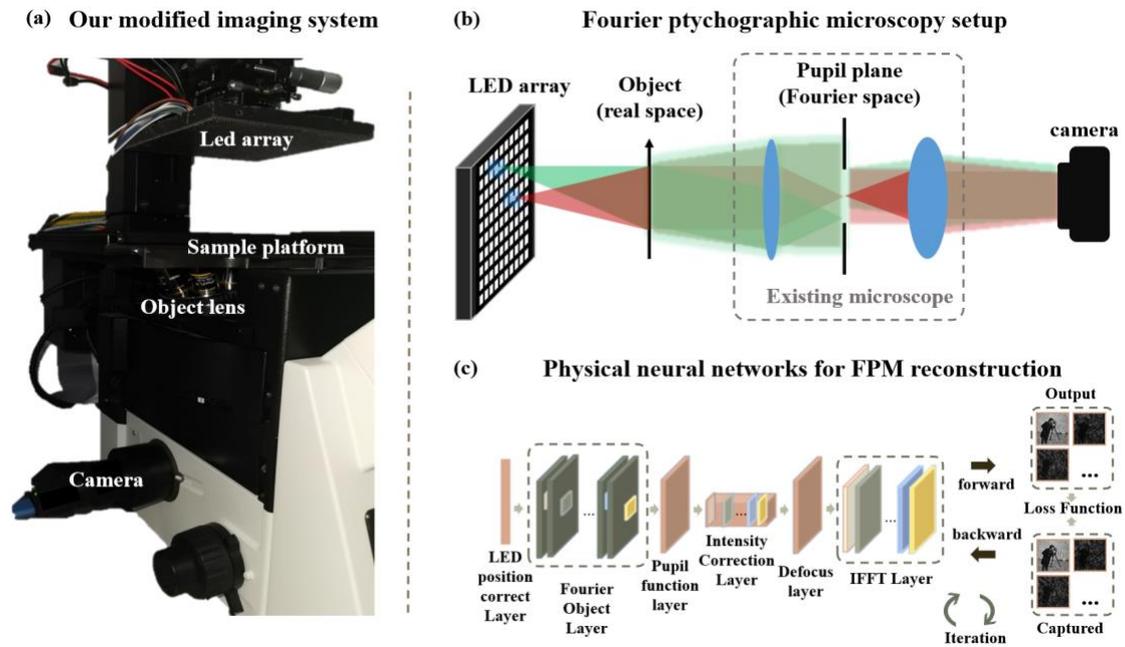

Fig. 2 FPM Experimental System. (a)The FPM system we modified on Olympus IX73. (b)FPM imaging system principle. (c)Physical neural network for image reconstruction with Stochastic Gradient Descent.

Fourier ptychographic microscopy (FPM) is a typical computational imaging technique with a high spatial bandwidth product (SBP) [24], garnering widespread attention in fields such as pathological diagnosis, materials science, and bioinformatics [25-28]. Unlike traditional whole slide imaging (WSI) systems [29], which rely on precise mechanical guides for spatial domain scanning and stitching, FPM operates in the frequency domain. Setting up a classic FPM system involves simply replacing the conventional microscope's light source with a programmable LED array. As shown in Fig. 2(a), we modified an Olympus IX73 to build our FPM imaging system, and Fig. 2(b) describes the data acquisition process, demonstrating the simplicity and ease of operation. Leveraging existing calibration and reconstruction algorithms [30-31], we believe that any pathologist with a spirit of exploration can modify their microscope into an FPM system for imaging.

In a typical FPM setup, different LED positions are illuminated to capture low-resolution (LR) images at various angles, achieving spectral scanning of the sample. Unlike traditional imaging methods that produce images directly, FPM models the imaging process and uses nonlinear iterative algorithms, involving multiple cycles of Fourier transforms and inverse transforms, to gradually recover the sample's complex amplitude. This allows for high SBP imaging without mechanical scanning, simultaneously obtaining high-resolution intensity and phase information. In FPM, the ideal forward propagation of the sample can be expressed by Eq. (1), where $I$ represents the captured LR image, $F$ denotes the two-dimensional Fourier transform, $k_m = (\sin\theta_{xm}/\lambda, \sin\theta_{ym}/\lambda)$ represents the spectral shift, $\lambda$ denotes the incident light wavelength, $(\theta_{xm}, \theta_{ym})$ denotes the tilt angle of the incident light, and $P(k)$ denotes the pupil function. In Eq. (2), $o(r)$ represents the object's spectrum, $r = (x, y)$ represents the sample layer coordinates, $\varphi(r)$ represents the phase modulation distribution, and $\mu(r)$ represents the absorption distribution.

In practical pathological diagnosis, accurate focusing of the sample is usually necessary for optimal observation. To simplify this labor-intensive process, a defocus term is introduced into the ideal propagation process. Eq. (1) is rewritten as Eq. (3), where $H(k,z) = \exp(j\frac{2\pi}{\lambda} \cdot z \cdot \sqrt{1 - k_x^2 - k_y^2})$ models the effect of defocus distance, and $\gamma_i$ is a real number representing the difference in luminous intensity.

$$o(r) = \exp(i\varphi(r) - \mu(r)) \qquad (1)$$

$$I_i = |F^{-1}(o(k - k_{mi})P(k))|^2, i = 1,2,\dots,n \qquad (2)$$

$$I_i = \gamma_i |F^{-1}(o(k - k_{mi})P(k)H(k,z))|^2, i = 1,2,\dots,n \qquad (3)$$

Building on previous work [31-32], we deployed a multi-parameter physical neural network for image reconstruction, as shown in Fig. 2(c). Unlike earlier methods [33-34], the physical neural network achieves multi-parameter joint optimization, effectively enhancing the reconstruction speed and quality of FPM, making it more suitable for clinical applications. We used L1 loss as the optimization target and updated the network parameters using the Adam optimizer [35]. Specifically, we used the low-resolution images captured by the central light as the initial values for the sample's recovered amplitude, with the initial defocus distance set to 0. The pupil function was also initialized as a Zernike polynomial, with only the first polynomial coefficient set to 1 and all other coefficients set to 0. We followed the principles outlined in [31] and, based on batch-FPM [40], deployed the entire algorithm on a GPU to accelerate image reconstruction. The reconstruction process of the physical neural network can be represented by minimizing Eq. (4), where $\|.\|^2$ denotes the Euclidean distance, N denotes the number of captured LR images, $I_{pred_i}$ denotes the predicted value of the sample spectrum, and $I_{gt_i}$ denotes the measured value captured by the camera.

$$\min \varepsilon = \frac{1}{n}\sum_{i=1}^{N}\sqrt{\|I_{pred_i} - I_{gt_i}\|^2} \qquad (4)$$

In actual experiments, we observed H&E-stained colon cancer sections using an image sensor with a pixel size of 6.5 μm and a 10X objective lens with an NA of 0.3. A 13 × 13 LED array illuminated the sample at a distance of 52 mm. Using illumination wavelengths of 523 nm, 470 nm, and 623 nm, we reconstructed the captured raw data on a workstation equipped with an RTX 3090 graphics card with 24GB of memory and a Silver 4210R CPU. We stitched the three reconstructed color results at the channel dimension to obtain an RGB image. This approach provides a robust solution for high-quality imaging and efficient reconstruction in FPM, with significant potential for applications in clinical diagnosis and research.

**3. Results**

3.1 Results statistics

Tumor budding images annotated using conventional microscopy and Fourier ptychographic microscopy (FPM) were compared to determine the budding counts within selected hotspot areas at 20x magnification (field of view: 0.785 mm² ). The locations and counts of the budding sites in both images were assessed for consistency. Table 1 shows a comparison of the counts of tumor buddings seen by conventional microscope and FPM on the same pathology section. The purpose of this comparison is to demonstrate whether the images obtained by the FPM, with its inherent advantage of not having to switch between high and low magnification and refocus, are able to accurately mark the location of tumor budding and whether the final counts of tumor budding are in agreement with those of the conventional microscope, and therefore to demonstrate that the FPM can be as accurate as the conventional microscope in the diagnosis of tumor budding. This demonstrates that FPM is as accurate as conventional microscopy for the diagnosis of tumor buddings while saving labour. It is illustrated that FPM can be well applied in the diagnosis of interpreting tumor budding. Table 1 presents the results from 10 randomly selected pathological sections out of 600, comparing the diagnostic outcomes of the two microscopy techniques. Fig.3 shows the labelling of tumor budding under the 20x objective of conventional microscopy and the labelling of tumour budding obtained by FPM at the 10x objective, which is equivalent to the resolution of a 20x objective in conventional microscopy. This illustrates the advantages of FPM in terms of its large field of view and high resolution, which are conducive to the rapid diagnosis of tumour budding.

Table1.Two kinds of microscopes were used to randomly select 10 tumor sprouts from 600 labeled tumor sprouts to show the specific counting results

| Colorectal cancer | Tumor budding count by conventional microscopy | Tumor budding count by FPM |
|---|---|---|
| 1 | 10 | 10 |
| 2 | 8 | 8 |
| 3 | 8 | 8 |
| 4 | 3 | 3 |
| 5 | 6 | 6 |
| 6 | 4 | 4 |
| 7 | 9 | 9 |
| 8 | 5 | 5 |
| 9 | 7 | 7 |
| 10 | 3 | 3 |

Display of the specific counting results of 600 cases randomly selected by pathologists with microscope and FPM labeling respectively It shows that the results of the two microscopes are consistent.

There were 600 pathological sections pathological in total. Using two kinds of microscopes to count the germination of tumors on pathological sections.Conventional microscopy method was considered as the ground truth. A specialist could count tumor budding using FPM with 100% accuracy.

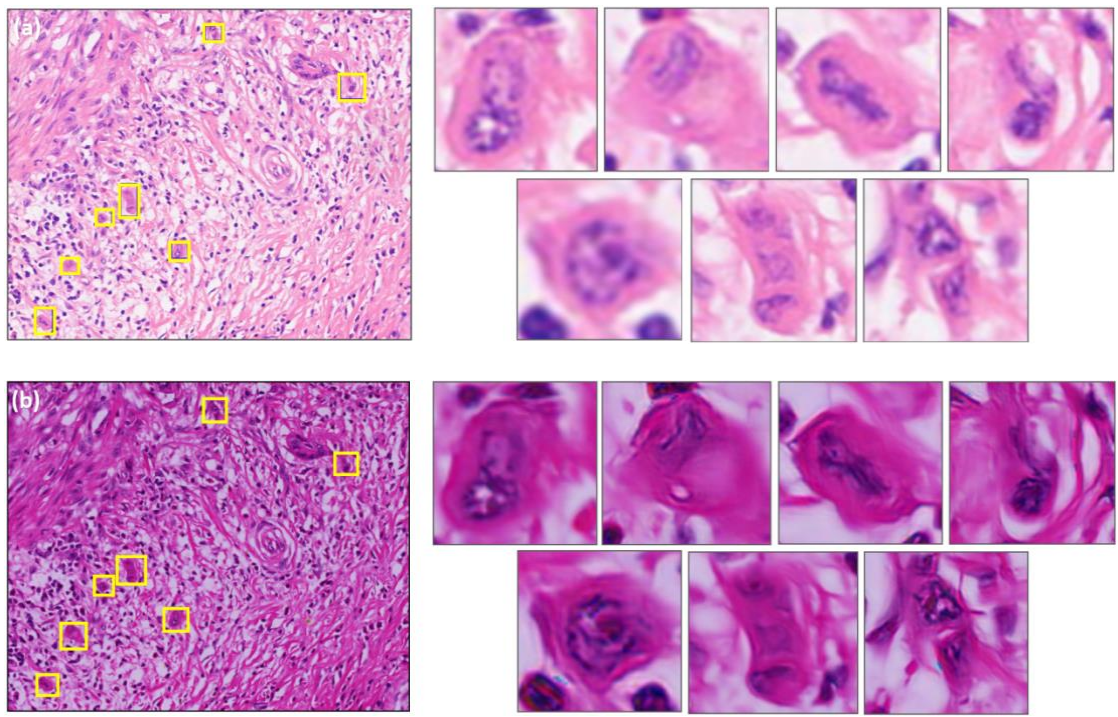

Fig. 3 (a)is a 20x objective image of an ordinary conventional microscope. Magnification of tumour budding and each budding localisation marked on (a). (b) is an image obtained from an 10x objective of FPM, with a resolution equivalent to that of a 20x objective in a conventional microscope. Magnification of the tumor budding and each outgrowth localisation marked on (b).

3.2  Focusing

Pathology slides are designed to be monofocal, but there is unfocusing due to the fact that tissues are being sliced, and most of the pathology slides vary in thickness during the preparation process. Unfocused areas can hinder the pathologist's ability to accurately observe and interpret tissue sections, requiring frequent manual adjustments between positions and magnifications, which increases workload and can lead to eye strain, especially with the growing number of patients. Instead of manually adjusting the focus, the pathology slides obtained by FPM can be automatically focused on the multifocal image as shown in Fig.4. Fig. 4(a) shows how a blurred and distorted image can mislead the pathologist, while Fig. 4(c) illustrates a fully focused multifocal image. Therefore, the application of FPM is both labour-saving and fast in obtaining high-resolution monofocal images, which provides a powerful help to pathologists in their clinical work.

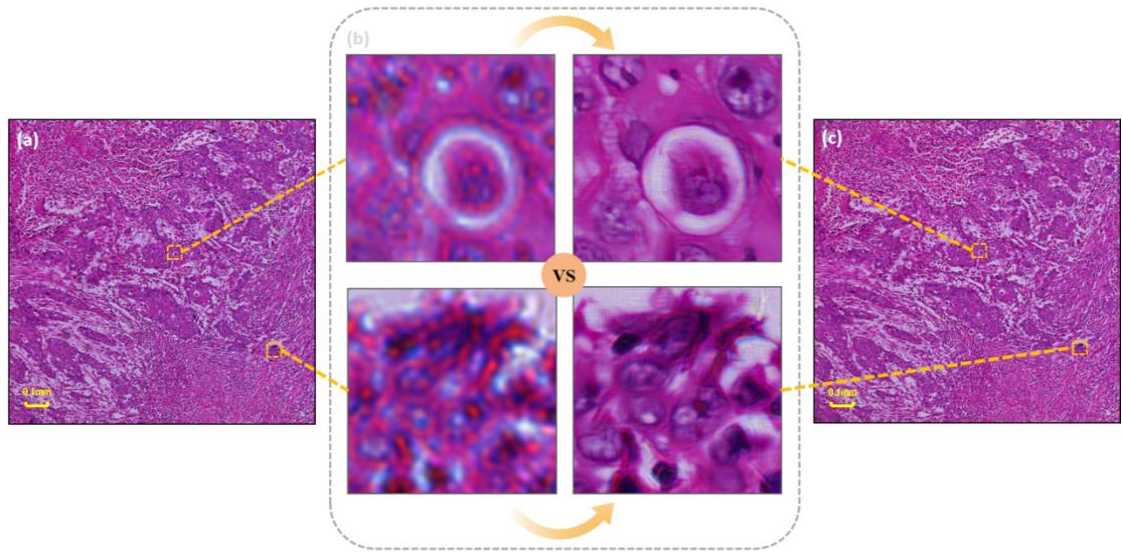

Fig. 4 (a)is a 10x objective image observed without manual focusing, and c is an autofocus 10x objective image obtained at FPM. (b)is a comparative display of 20x objective images of part of the field of view of (a) and (c).

3.3  Mark removal

Due to human or external factors of contamination, will lead to some pathology slides on the existence of certain disturbing dust, stains, water stains, scratches, handprints and markers mark to a certain extent will affect the pathologist's judgement of the tumor cells. FPM can also solve the problem, to get a large field of view of high-resolution and clear images.If HE slides with markings are directly placed under the microscope for observation, those markings will obscure the normal tissue cells. As a result, pathologists must clean the slides before re-observing them. FPM eliminates the cleaning process for pathologists on pathology slides. FPM allows slides to be placed directly under the microscope, even with surface interference. Using the FPM algorithm for reconstruction, as shown in Fig. 5, clear, high-resolution images are obtained. This not only saves the pathologist the time and effort of cleaning but also yields clear, high-resolution images.

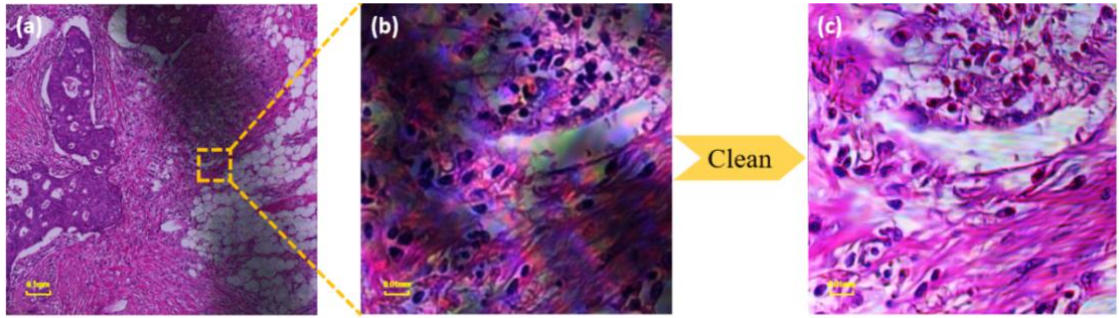

Fig.5 (a)shows a pathological image with marker pen marks,(b)is a partial enlargement of the stained area.The stain can be removed automatically under FPM to get the effect of (c).

3.4  quantitative phase imaging

FPM acquired the H&E images along with the corresponding quantitative phase maps, which acquired the entire field of view of the temporal phase maps of intestinal tissues, and the overall morphological structure of intestinal tissues could be obtained in the phase maps. At the same time, the microscopic details of cancer tissues at the level of individual cancer cells were investigated. We compared the H&E image in Fig. 6(a) obtained by FPM with the phase diagram in Fig. 6(c), both of which clearly show the morphological details of intestinal tissue. The nuclei are relatively obvious in the phase diagram because the presence of HE staining alters the local light absorption properties of this tissue structure. Tumor tissue and fibrous connective tissue can also be identified in Figs. 6(b) and 6(c), offering clear guidance to the pathologist during diagnosis.

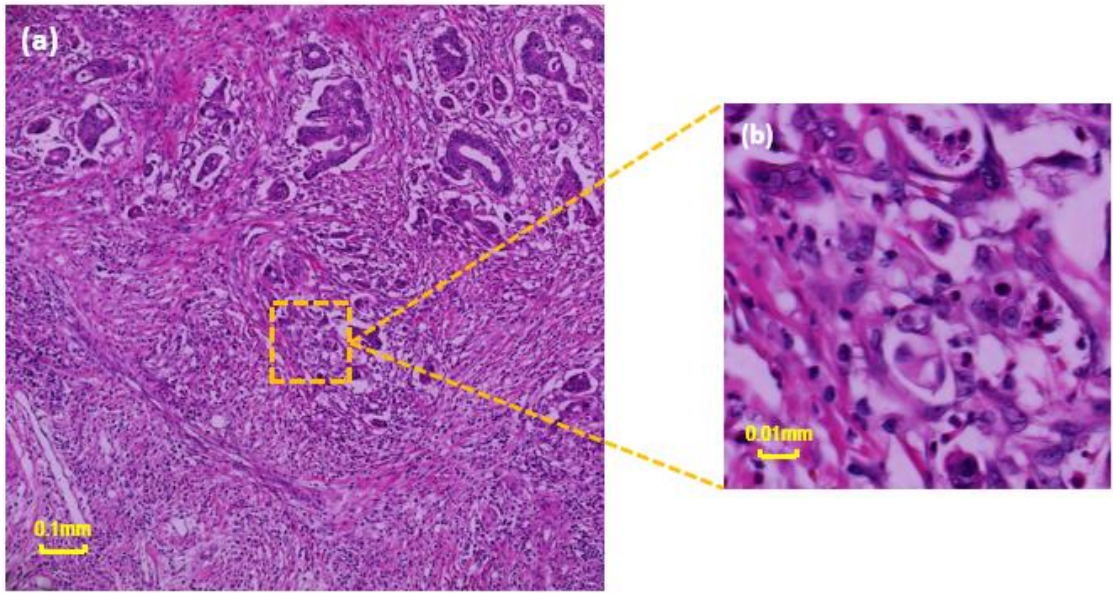

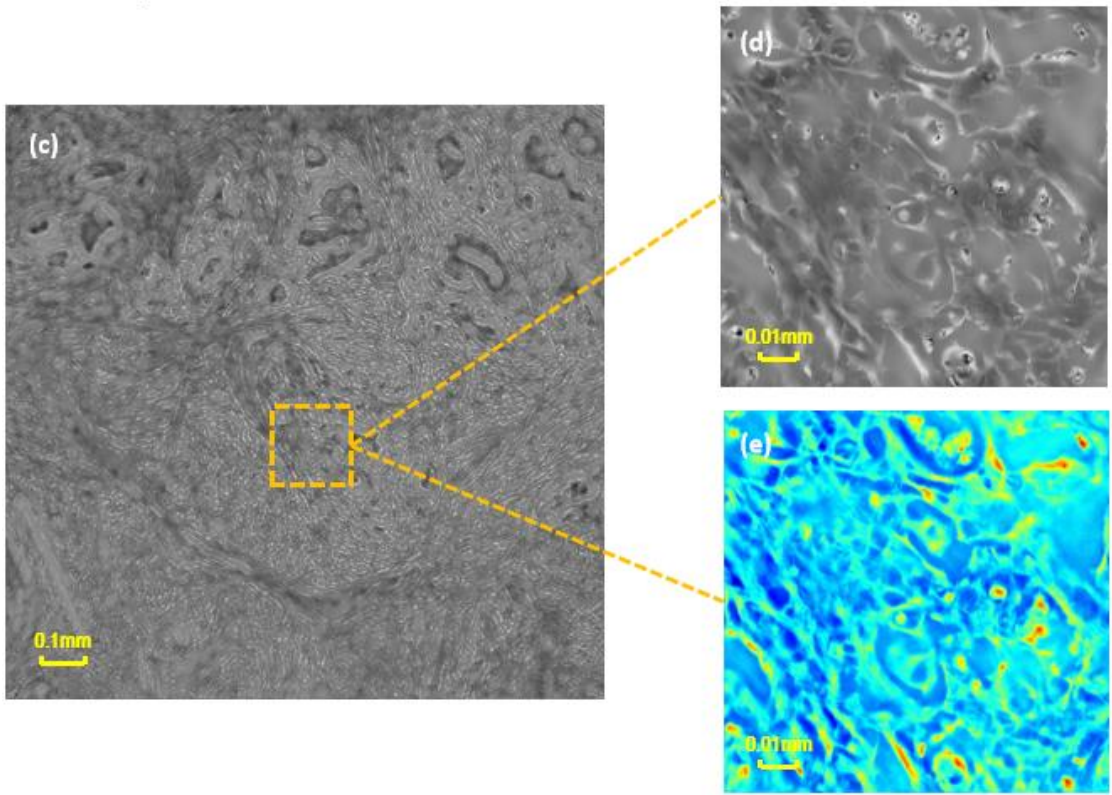

Fig.6 (a)is the HE image obtained by FPM, (b)is the local zoomed-in image. (c)is the phase map obtained simultaneously by FPM, (d)and (e)are the local zoomed-in images of the same region.

**4    Discussion and conclusion**

Currently, the use of FPM is a good alternative for disease diagnosis of tumor budding, FPM is a technique with the advantages of high resolution, large field of view, autofocus and quantitative phase recovery [18-20] [36-37].Our study demonstrates that when observing colorectal cancer tumor budding with FPM, tumor budding that is completely consistent with traditional optical microscopy can be accurately marked without high and low magnification switching and repeated focusing. At the same time, the quantitative phase image obtained by FPM can more accurately observe the nucleus, tumor tissue and fibrous connective tissue. All of these provide a certain direction for pathologists to correctly interpret.The FPM's autofocus is also highly effective, solving the problem of multiple focal surfaces on pathology slides. Automatic stain removal is also effective, reducing the pathologist's labour while producing clear pathology images. All these advantages compared with the traditional optical microscope only need to add a programmable LED light-emitting diode, which is a simple and costless modification, providing great convenience to pathologists [38-39].We can also use the information provided by quantitative phasing to add to deep or machine learning, thus increasing the accuracy of the training set so that tumor budding can be identified more accurately and automatically.

  We propose that FPM be heavily used in clinical practice as an alternative imaging tool.At the same time, it is necessary for us to carry out further research to explore the potential ability of FPM in clinical pathological applications and improve the quantitative phase effect map of FPM. We hope that FPM can combine the extracted feature information with clinicopathological features to distinguish between healthy and cancerous tissues in colorectal cancer. The ultimate goal is to help pathologists make more accurate and rapid diagnoses.

**Declarations**

**Ethics approval** The use of materials and clinical information was approved by the Research Ethics Committee of Shanxi Cancer Hospital(reference code KY2023005).

**Conflicts of interest** All authors declare that there is no conflicts of interest

**Funding** This work received funding support from the National Natural Science Foundation of China (Grant Number: 82172659), "Four Batches" of the Science and Technology Innovation Plan of Shanxi Provincial Health Commission of China (Grant Number: 2020TD07), The Science and Education Cultivation Fund of the National Cancer and Regional Medical Center of Shanxi Provincial Cancer Hospital （Grant Number: BD2023006）.

**Authors' contributions**

Methodology:

Writing – original draft: Yiyan Su, Ruiqing Sun

Writing – review & editing: Yajing Wang, Yanfeng xi, Shaohui Zhang

**Availability of data and materials** All data generated or analyzed during this study are included in this published article

**Consent for publication** The participants provided informed consent for the publication of the study.

**Competing interests** The author(s) declare that they have no conflict of interests.

**Data openly available in a public repository.**